\documentclass[preprint2]{aastex}

\begin{document}


\title{Reply to ``Comment to a paper of M.\ Villata on antigravity"}


\author{M. Villata}
\affil{INAF, Osservatorio Astronomico di Torino, Via Osservatorio 20, I-10025 Pino Torinese (TO), Italy}
\email{villata@oato.inaf.it}


\begin{abstract}
In this short paper we reply to the Comment by M.\ J.\ T.\ F.\ Cabbolet on Villata's theory of antigravity. The criticisms of methodological and ontological kind presented by that author come from a misinterpretation of some concepts, perhaps due to some lack of clarity or omission of details in Villata's original article. In order to clarify these points, here we provide additional explanations regarding the assumptions and results of the theory.
\end{abstract}






In his Comment \citep{cab11}, the author criticizes Villata's theory of antigravity from the methodological and ontological points of view. But, some concepts are misinterpreted by him, and, consequently, also his ontological discussion is not appropriate.

In his paper, \citet{vil11} starts from the well-established general theory of relativity, and extends it in the most natural way to incorporate antimatter. This is done in agreement with the Feynman-Stueckelberg interpretation for antimatter as matter that travels backwards in time.

Indeed, there are two possible interpretations for the existence of antimatter. The more conventional one is that antiparticles really exist as entities distinct from their matter counterparts, and that they travel forward in time (FIT), as all ordinary particles. The other interpretation is that antiparticles do not really exist as distinct particles, but that they are nothing else than the corresponding particles that are traveling backwards in time (BIT). There are several reasons why the latter appears more convincing.

In the BIT interpretation the number of existing particle species is halved and time is symmetrized. It may appear as a matter of taste, but the need of doubling particle species with quasi-identical (but marginally existing) ones looks like an artifact. Even more strange is the unusual tendency of particles and antiparticles to destroy each other (or to be created in pairs). More interesting and elegant is the view of a particle changing its time direction by emitting or absorbing photons. But the most convincing argument in favor of the BIT interpretation is just the CPT symmetry of physical laws, since it offers a physical explanation to the need of coupling C with T (and P) for describing the behavior of antimatter: if antimatter were not traveling back in time, why should we apply the time inversion? In other words, according to the BIT interpretation, CPT is the operation that transforms events, particles and fields from one time direction to the other: the role of T is obvious, since it inverts time intervals, P is needed to get a proper Lorentz transformation (T and P alone are improper), and C provides the needed charge reversals to see time-reversed matter as antimatter.

In any case, FIT or BIT interpretation, CPT symmetry is universal and assured by the CPT theorem, while the C-only symmetry usually adopted to transform matter into antimatter is not guaranteed (and indeed it is violated in weak interactions), and may be seen as a deceptive effect of theories that are C, P and T invariant also separately.

These are the reasons why the CPT transformation has to be considered as the correct operation to convert matter into antimatter in a given physical system governed by CPT-invariant laws, instead of the more conservative C-only prescription.

The first flaw in Cabbolet's Comment is his criticism, in the Introduction, that CPT symmetry, coming from another theory, should not be taken as an assumption in general relativity. However, as it is well known, the CPT theorem is well established in flat space-time, and, even if it is not proven in curved space-time, nobody expects that general relativity violates it (apart from the case of very peculiar topologies). Thus it is not surprising that Villata first assumes CPT invariance and then checks it in the relevant gravitational equations: as expected, since the C operation is ineffective there, the well-known PT invariance is verified.

Moreover, Cabbolet uses some incorrect expressions, such as ``a new equation is constructed" and ``This is then interpreted as the equation that\dots". Actually, eq.~(9) is not a new equation that is first constructed and then interpreted, but it is the direct transformation of the equation of motion of general relativity extended to consider the gravitational interaction between objects belonging to the two opposite space-time reference frames, i.e.\ between matter and antimatter.

In Sect.~2 of \citet{cab11}, the author speaks of ``cart before the horse" and ``inadmissible method", and here he is partly right, in the sense that in his paper Villata is more concerned with scientific results and to give an intuitive and understandable view of the theory rather than to follow a rigorous and unassailable (but less effective) methodology. Now we hope that what has been explained above can clarify the matter and fill that gap.

Regarding the discussion on the ontology of general relativity and the thought experiment proposed by Cabbolet, there is a bit of confusion.

Unlike what is affirmed in \citet{cab11}, the gravitational field is not the metric of space-time, but some combination of derivatives of the metric tensor, and, while the metric tensor is a rank-2 tensor and hence is (C)PT-even, the gravitational field is (C)PT-odd and so it is ``seen" inverted from the reversed space-time of the antineutron, or, from the ``point of view" of the gravitational field space-time, the antineutron is ``seen" with an inverted gravitational charge. The confusion in the discussion by Cabbolet is perhaps due to the fact that we speak of ``opposite space-times", but everything is observed and measured from a single space-time, the observer one, while the other is treated like any other reference frame, and the (C)PT transformation is what translates what happens in the other frame into what is observed. Thus, Villata's theory is merely the extension of general relativity to consider the opposite space-time, which is observed through a simple Lorentz transformation, PT, so that it is not in conflict with the ontological presuppositions of general relativity. It would be different if we wanted to construct a theory that ``wander" from one space-time to the other, e.g.\ following the neutron along its world line in the BIT interpretation, which, in the case of the neutron-antineutron pair production, would arrive from our future, would reverse its time direction, and would return (with us) towards the future: then yes that a single space-time would no longer be the set for all events.

Consequently, the discussion and conclusions in Sect.~3 in \citet{cab11} are not appropriate. While Santilli's and Cabbolet's theories require new mathematical and ontological approaches, Villata's theory is the mere and immediate extension of the well-established general theory of relativity to the already existing interpretation that antimatter is matter that goes backwards in time, without any other new assumption or construction of new formalism. In other words, antigravity comes out as a result of already existing theories.

Even though Villata's theory is not concerned with the unification of general relativity with quantum mechanics, Cabbolet's statement that ``it will thus not bring a solution to this main problem of contemporary physics any closer" should be demonstrated. Indeed, merely the fact that antigravity is predicted is in the right direction to find the interface between the two theories.

Finally, the author quotes the famous argument against antigravity: ``a photon is identical to its antiparticle, yet it sees the same metric as ordinary matter".

The explanation is simple. In general relativity, the electric charge and other quantum properties that distinguish particles from their antiparticles do not affect gravity. In other words, C is ineffective anyway, no matter whether particle and antiparticle are identical or different from each other. What works is PT, which converts retarded photons into advanced photons, i.e.\ those going back in time (as expected for photons emitted by antimatter). And these advanced photons would be repelled by a matter gravitational field. Indeed, the equation of motion for photons is formally identical to that for material particles:
\begin{equation}
\label{eq.1}
{{\rm d}^2x^\lambda\over{\rm d}\sigma^2}=-\Gamma^{\lambda}_{\mu\nu}{{\rm d}x^\mu\over{\rm d}\sigma}{{\rm d}x^\nu\over{\rm d}\sigma}\,,
\end{equation}
where $\sigma$ is a parameter describing the trajectory, which for photons can not be equal to the proper time $\tau$, since ${\rm d}\tau=0$, and its elements have the same (C)PT properties as in the equation for massive particles.

But advanced photons can not be observed (at least not at the astrophysical-cosmological level and with the usual detectors), since from our opposite time direction the roles of emitter and absorber are interchanged, and then they are destined to ``return" to their source: if we could detect them, i.e.\ intercept them along this path, they would never have been emitted, i.e.\ they could not exist, even though this may sound like a contradiction.

Similarly, visible photons, i.e.\ retarded ones, are repelled by an antimatter gravitational field, thus providing a test for the theory: even if antimatter can not be detected directly by its advanced photons, it could be revealed by its gravitational effect on the retarded radiation coming from the background, giving rise to an antigravitational lensing, i.e.\ a diverging lens type effect, which brings the images of the background sources closer to the lensing object.

It could thus seem that matter emits only retarded radiation and antimatter only advanced radiation, and this would provide an immediate explanation for the invisibility of antimatter in the Universe, in particular regarding its presence in cosmic voids suggested in \citet{vil11}. But this hypothesis is not necessarily true. If possible antimatter stars and galaxies emit also retarded radiation (advanced from their point of view), then they could be well visible. In this case, the invisibility of antimatter in cosmic voids must have another explanation: either it is in a state that does not emit at all, or the emission is weak or strongly absorbed somewhere. But a more thorough discussion on this topic is beyond the aim of this paper.

Some of the arguments discussed in this paper also provide answers to the objections raised by \citet{cro11} in his \emph{Response to ``CPT symmetry and antimatter gravity in general relativity"\/}.




\end{document}